\documentclass[final]{IEEEtran}

\usepackage{cite}
\usepackage{threeparttable}
\usepackage{graphicx}
\usepackage{picinpar}
\usepackage{hyperref}
\usepackage[cmex10]{amsmath}
\usepackage{amsmath,amsfonts,amssymb}
\usepackage{subfigure}

\usepackage{algorithm}
\usepackage{algorithmic}
\usepackage{stfloats}
\usepackage{bm}
\usepackage{setspace}

\begin{document}
\newtheorem{lemma}{Lemma}
\newtheorem{corol}{Corollary}
\newtheorem{theorem}{Theorem}
\newtheorem{proposition}{Proposition}
\newtheorem{definition}{Definition}
\newcommand{\e}{\begin{equation}}
\newcommand{\ee}{\end{equation}}
\newcommand{\eqn}{\begin{eqnarray}}
\newcommand{\eeqn}{\end{eqnarray}}

\title{Near-Optimal Signal Detector Based on Structured Compressive Sensing for Massive SM-MIMO}
\author{

Zhen Gao, Linglong Dai, Chenhao Qi, Chau Yuen, and Zhaocheng Wang
\thanks{Copyright (c) 2015 IEEE. Personal use of this material is permitted. However, permission to use this material for any other purposes must be obtained from the IEEE by sending a request to pubs-permissions@ieee.org. This work was supported by the National Key Basic Research Program of China (Grant No. 2013CB329203), the National Natural Science Foundation of China (Grant Nos. 61571270, 61302097, and 61271266), the Beijing Natural Science Foundation (Grant No. 4142027), and the Foundation of Shenzhen government.}
\thanks{Z. Gao, L. Dai, and Z. Wang are with Tsinghua National Laboratory for
 Information Science and Technology (TNList), Department of Electronic Engineering,
 Tsinghua University, Beijing 100084, China (E-mails: gaozhen010375@foxmail.com; \{daill,zcwang\}@mail.tsinghua.edu.cn).}
 \thanks{C. Qi is with School of Information Science and Engineering, Southeast University, Nanjing 210096, China (E-mail: qch@seu.edu.cn).}
 \thanks{C. Yuen is with Singapore University of Technology and Design, Singapore 138682, Singapore (E-mail:~yuenchau@sutd.edu.sg).}
   }
\maketitle
\begin{abstract}
Massive spatial modulation (SM)-MIMO, which employs massive low-cost antennas but few power-hungry transmit radio frequency (RF) chains at the transmitter, is recently proposed to provide both high spectrum efficiency and energy efficiency for future green communications. However, in massive
SM-MIMO, the optimal maximum likelihood (ML) detector has the prohibitively high complexity, while
state-of-the-art low-complexity detectors for conventional small-scale SM-MIMO suffer from an obvious performance loss. In this paper, by exploiting the structured sparsity of multiple SM signals, we propose a low-complexity signal detector based on structured compressive sensing (SCS) to improve the signal detection performance.
Specifically, we first propose the grouped transmission scheme at the transmitter, where multiple SM
signals in several continuous time slots are grouped to carry the common spatial constellation
symbol to introduce the desired structured sparsity. Accordingly, a structured subspace pursuit (SSP)
algorithm is proposed at the receiver to jointly detect multiple SM signals by leveraging the structured sparsity. In addition, we also propose the SM signal interleaving to permute SM signals
in the same transmission group, whereby the channel diversity can be exploited to further improve the signal detection performance.
Theoretical analysis quantifies the performance gain from SM signal interleaving, and simulation results demonstrate the near-optimal performance of the proposed scheme.
\end{abstract}
\section{Introduction}
Spatial modulation (SM)-MIMO exploits the pattern of one or several simultaneously active antennas out of all available transmit antennas
to transmit extra information~\cite{LS_SM}, \cite{design_guide}. Compared with small-scale SM-MIMO which only introduces the limited gain in spectrum efficiency, 
massive SM-MIMO is recently proposed by integrating SM-MIMO with massive MIMO working at 3$\sim$6 GHz to achieve higher spectrum efficiency~\cite{LS_SM}. In massive SM-MIMO systems, the base station (BS) uses a large number of low-cost antennas for higher spectrum efficiency
but only one or several power-hungry transmit radio frequency (RF) chains for power saving, while the user can compactly employ the multiple receive diversity antennas with low correlation\cite{design_guide}.
Since the power consumption and hardware cost are largely dependent on the number of simultaneously active transmit RF chains (especially the power amplifier), massive SM-MIMO outperforms the traditional MIMO schemes in higher spectrum efficiency, reduced power consumption, lower hardware cost, etc.
In practice, SM can be adopted in conventional massive MIMO systems as an energy-efficient transmission mode. Meanwhile, massive SM-MIMO can be also considered as an independent scheme to reduce both the power consumption and hardware cost.
To date, in addition to the combination of massive MIMO and SM, the concept of SM has also been integrated into various applications \cite{design_guide} including cognitive radio \cite{spectrum_sharing,spectrum_sharing2}, physical-layer security~\cite{secret}.

SM-MIMO maps a block of information bits into two information carrying units: the spatial constellation symbol
 and the signal constellation symbol.
For massive SM-MIMO, due to the small number of receive antennas at the user and massive antennas at the BS, the signal detection is a challenging large-scale underdetermined problem. When the number of transmit antennas becomes large, the optimal maximum likelihood (ML) signal detector suffers from the prohibitively high complexity \cite{SVD}.
Low-complexity signal vector (SV)-based detector has been proposed for SM-MIMO \cite{SVD}, but it is confined to SM-MIMO with single transmit RF chain.
In \cite{{ML1},{ML2},{WJT}}, the SM is generalized, where more than one active antennas are used to transmit independent signal constellation symbols for spatial multiplexing.
Linear minimum mean square error (LMMSE)-based signal detector \cite{LS_SM} and sphere decoding (SD)-based detector \cite{SD} can be used for SM-MIMO systems with multiple transmit RF chains. But they are only suitable for well or overdetermined SM-MIMO with $N_r \ge N_t$, and suffer from a significant performance loss in underdetermined SM-MIMO systems with $N_r < N_t$, where ${N_t}$ and ${N_r}$ are the numbers of transmit and receive antennas, respectively. 
%
Due to the fact that the number of active antennas is smaller than the total number of transmit antennas, SM signals have the inherent sparsity, which can be considered by exploiting the compressive sensing (CS) theory \cite{{STR_CS},{SP}} for improved signal detection performance. By far, CS has been widely used in wireless communications \cite{{shim},{uplink},{CS_CL1},{CS_CL2}}, and the CS-based signal detectors have been proposed for underdetermined small-scale SM-MIMO \cite{{CS_CL1},{CS_CL2}}. However, their bit-error-rate (BER) performance still has a significant gap compared with that of the optimal ML detector, especially in massive SM-MIMO with large $N_t$, $N_r$, and $N_r \ll N_t$.

To this end, this paper proposes a near-optimal structured compressive sensing (SCS)-based signal detector with low complexity for massive SM-MIMO. 
Specifically, we first propose the grouped transmission scheme at the BS, where multiple SM signals in several successive time slots are grouped to carry the common spatial constellation symbol to introduce the desired structured sparsity. 
Accordingly, we propose a structured subspace pursuit (SSP) algorithm at the user to detect multiple SM signals in the same transmission group, whereby their structured sparsity is leveraged for improved signal detection performance. Moreover, 
the SM signal interleaving is proposed to permute SM signals in the same transmission group, so that the channel diversity can be exploited. 
Theoretical analysis and simulation results verify that the proposed SCS-based signal detector outperforms existing CS-based signal detector.



\textit{Notation}: 
 Boldface lower and upper-case symbols represent column
 vectors and matrices, respectively. $\lfloor \cdot \rfloor$ denotes the integer floor operator. The transpose, conjugate transpose, and Moore-Penrose matrix inversion operations are denoted by $(\cdot )^{ T}$, $(\cdot )^{ *}$ and $(\cdot )^{\dag}$, respectively. The $l_p$ norm operation is given by $\|\cdot\|_p$, and $| \cdot |$ denotes the cardinality of a set.
$E\left\{ {\cdot} \right\}$, ${\mathop{\rm var}} \left\{ {\cdot} \right\}$, ${\mathop{\rm Re}\nolimits} \left\{ {\cdot} \right\}$, and ${\mathop{\rm Im}\nolimits} \left\{ {\cdot} \right\}$ are operators to take the expectation, variance, the real part, and the imaginary part of a random variable. ${\rm{Tr}}\left\{ {\cdot} \right\}$ is the trace operation for a matrix.
 If a set has $n$ elements, the number of $k$-combinations is denoted by the binomial coefficient $\binom{n}{k}$.
 The index set of non-zero entries of the vector $\mathbf{x}$ is called the support set of $\mathbf{x}$, which is denoted by ${\rm supp} \{\mathbf{x}\}$, $ \mathbf{x}_{i}$ denotes the $i$th entry of the vector $\mathbf{x}$, and $ \mathbf{H}_{i}$ denotes the $i$th column vector of the matrix $\mathbf{{H}}$. $ \mathbf{x}_{\Gamma}$ denotes the entries of $\mathbf{x}$ defined
 in the set $\Gamma$, while $\mathbf{H}_{\Gamma}$ denotes a sub-matrix of $\mathbf{H}$ with indices of columns defined by the set~$\Gamma$.
\section{System Model}\label{model}

\begin{figure}[!tp]
     \centering
     \includegraphics[width=9.0cm, keepaspectratio]
     {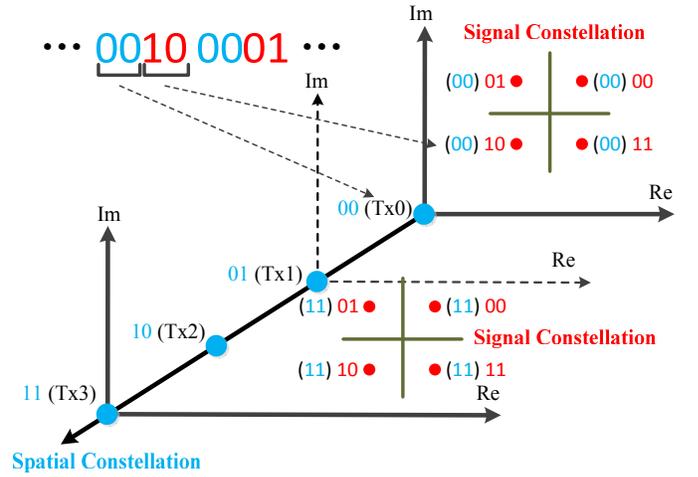}
    \caption{Spatial constellation symbol and signal constellation symbol in SM-MIMO systems, where $N_t=4$, $N_a=1$, and QPSK are considered as an example.} 
     \label{3D}
\end{figure}

The SM-MIMO systems can be illustrated in Fig. \ref{3D}. The transmitter has $N_t$ transmit antennas but $N_a<N_t$ transmit RF chains, and the receiver has $N_r$ receive antennas. Each SM signal consists of two symbols: the spatial constellation symbol obtained by mapping $\left\lfloor {{\rm{log}}_{2}\binom{N_t}{N_a}} \right\rfloor $ bits to a pattern of $N_a$ active antennas out of $N_t$ transmit antennas, and ${N_a}$ independent signal constellation symbols coming from the $M$-ary signal constellation set (e.g., QAM).
Hence, each SM signal carries the information of ${N_a}{\log _2}M + \left\lfloor {{\rm{log}}_{2}\binom{N_t}{N_a}} \right\rfloor$~bits. 

At the receiver, the received signal ${\bf{y}}\in \mathbb{C}^{N_r  \times 1}$ can be expressed as
${\bf{y}} = {\bf{Hx}} + {\bf{w}}$, where ${\bf{x}}\in \mathbb{C}^{{N_t} \times 1}$ is the SM signal transmitted by the transmitter, ${\bf{w}}\in \mathbb{C}^{{N_r} \times 1}$ is the additive white Gaussian noise (AWGN) vector with independent and identically distributed (i.i.d.) entries following the circular symmetric complex Gaussian distribution ${\cal {CN}}(0,{\sigma ^2_w})$, ${\bf{H}}{\rm{ = }}{\bf{R}}_r^{1/2}{\bf{\tilde H}} {\bf{R}}_t^{1/2}\in \mathbb{C}^{{N_r} \times {N_t}}$ is the correlated flat Rayleigh-fading MIMO channel, entries of ${\bf{\tilde H}}$ are subjected to the i.i.d. distribution ${\cal {CN}}(0,1)$, ${\bf{R}}_r^{}$ and ${\bf{R}}_t^{}$ are the receiver and transmitter correlation matrices, respectively \cite{CE_SM}. The correlation matrix $\bf{R}$ is given by ${r_{ij}} = {r^{\left| {i - j} \right|}}$, where $r_{ij}$ is the $i$th row and $j$th column element of $\bf{R}$, and $r$ is the correlation coefficient of neighboring antennas.

It should be pointed out that since ${\bf{H}}$ is used for conveying information, it should be known by the receiver and can be acquired by channel estimation \cite{CE_SM}.
To achieve both high spectrum efficiency and energy efficiency for future green communications, massive SM-MIMO, which employs massive low-cost antennas but few power-hungry transmit RF chains at the BS to serve the user with comparatively small number of receive antennas, is recently proposed~\cite{LS_SM}. However, its signal detection is a challenging large-scale underdetermined problem, since $N_t$, $N_r$ can be large and $N_r \ll N_t$, e.g., $N_t=64$ and $N_r=16$ are considered~\cite{LS_SM}.

For the SM signal ${\bf{x}}$, the spatial constellation symbol of $\left\lfloor {{\rm{log}}_{2}\binom{N_t}{N_a}} \right\rfloor $ bits is mapped into the spatial constellation set $\mathbb{A}$, where the pattern of $N_a$ active antennas selected from $N_t$ transmit antennas is regarded as the spatial constellation symbol. Hence there are $\left| {\mathbb{A}} \right|= {2^{\left\lfloor {{\rm{log_2}}\binom{N_t}{N_a}} \right\rfloor }}$ kinds of patterns of active antennas, i.e., ${\rm{supp}}\left\{ {\bf{x}} \right\} \in \mathbb{A}$. Meanwhile, the signal constellation symbol of the $i$th active antenna, denoted as ${x^{(i)}}$ for $1\le i \le N_a$, is mapped into the $M$-ary signal constellation set $\mathbb{B}$. Therefore, the signal detection in SM-MIMO can be formulated as the $ {M^{{N_a}}}{2^{\left\lfloor {{\rm{log_2}}\binom{N_t}{N_a}} \right\rfloor }}$-hypothesis detection problem.
Clearly, the optimal signal detector to this problem is ML signal detector, which can be expressed as~\cite{LS_SM}
\begin{equation}
\begin{array}{l}
{{\bf{\hat x}}_{\rm{ML}}} = \arg {\kern 1pt} {\kern 1pt} \mathop {\min }\limits_{{\mathop{\rm supp}\nolimits} ({\bf{x}}) \in {\mathbb{A}} ,{x^{(i)}} \in {\mathbb{B}} ,1 \le i \le {N_a}} {\left\| {{\bf{y}} - {\bf{Hx}}} \right\|_2}.\label{equ:ML}
\end{array}
\end{equation}
However, the computational complexity of the optimal ML signal detector is ${\cal{O}}({M^{{N_a}}}{2^{\left\lfloor {{\rm{log_2}}\binom{N_t}{N_a}} \right\rfloor }})$, which can be unrealistic when $N_t$, $N_a$, and/or $M$ become large. 

To reduce the complexity, SV-based signal detector has been proposed \cite{SVD}, but it only considers the case of $N_a=1$. LMMSE-based signal detector with the complexity of ${\cal{O}}(2N_rN_t^2 +N_t^3)$ \cite{LS_SM} and SD-based signal detector with the complexity of ${\cal{O}}(\max \{ N_t^3,N_r N_t^2, N_r^2 N_t\} )$ \cite{SD} have been proposed for well or overdetermined SM-MIMO with $N_r \ge N_t$. However, for underdetermined SM-MIMO systems with $N_r < N_t$, these detectors suffer from a significant performance loss \cite{CS_CL1}. Since only $N_a$ transmit antennas are active in each time slot for power saving and low hardware cost, there are only $N_a < N_t$ nonzero entries in $\bf{x}$, and thus the SM signal has the inherent sparsity. By exploiting such sparsity, the CS-based signal detectors have been proposed for SM \cite{{uplink},{CS_CL1},{CS_CL2}}. \cite{{uplink}} proposed a spatial modulation matching pursuit (SMMP) algorithm to detect multi-user SM signals in the uplink massive SM-MIMO systems. In \cite{{CS_CL1},{CS_CL2}}, the CS-based signal detectors are proposed for underdetermined single-user SM-MIMO systems with $N_r <N_t$ in the downlink. The normalized compressive sensing (NCS) detector (with the complexity of ${\cal{O}}(2N_r N_a^2 + N_a^3)$) in \cite{{CS_CL2}} first normalizes the MIMO channels and then uses orthogonal matching pursuit (OMP) algorithm to detect signals. \cite{{CS_CL1}} developed a basis pursuit de-noising (BPDN) algorithm (with the complexity of ${\cal{O}}(N_t^3)$) from the classical basis pursuit (BP) algorithm to detect SM signals. However, both NCS and BPDN detectors are based on the framework of CS theory, and such CS-based signal detectors still suffer from a significant performance gap compared with the optimal ML detector when $N_t/N_r$ becomes large, especially in massive SM-MIMO systems with $N_r \ll N_t$~\cite{CS_CL1}. 




\section{Proposed SCS-Based Signal Detector}
In this section, an SCS-based signal detector is proposed for downlink single-user massive SM-MIMO, which can be illustrated in Fig. 2. We first propose a grouped transmission scheme and an SM signal interleaving at the transmitter. Then, the corresponding deinterleaving and SSP algorithm for signal detection at the receiver are provided, whereby multiple SM signals with the structured sparsity are jointly processed for the improved signal detection performance with low complexity. 
\subsection{Grouped Transmission and Interleaving at the Transmitter}
Similar to conventional SM-MIMO systems, we assume that signal constellation symbols in the proposed scheme are mutually independent. However, unlike conventional SM signals, where spatial constellation symbols are independent in different time slots, we propose the grouped transmission scheme at the transmitter, where every $G$ SM signals in $G$ consecutive time slots are considered as a group, and SM signals in the same transmission group share the same spatial constellation symbol, i.e.,
\begin{equation}
\begin{array}{l}
{\rm{supp}}\left( {{{\bf{x}}^{\left( 1 \right)}}} \right) = {\rm{supp}}\left( {{{\bf{x}}^{\left( 2 \right)}}} \right) =  \cdots  = {\rm{supp}}\left( {{{\bf{x}}^{\left( {{G}} \right)}}} \right),\label{commonsparse}
\end{array}
\end{equation}
where $ {{{\bf{x}}^{\left( 1 \right)}}} $, $ {{{\bf{x}}^{\left( 2 \right)}}} $, $ \cdots $,  $ {{{\bf{x}}^{\left( G \right)}}} $ are SM signals in $G$ consecutive time slots. Due to the conveyed common spatial constellation symbol, $ {{{\bf{x}}^{\left( 1 \right)}}} $, $ {{{\bf{x}}^{\left( 2 \right)}}} $, $ \cdots $,  $ {{{\bf{x}}^{\left( G \right)}}} $ in the same transmission group share the same support set and thus have the structured sparsity. It is clear that to introduce such structured sparsity, the effective information bits carried by spatial constellation symbols will be reduced. However, as will be discussed in Section IV-C and demonstrated in our simulations, such structured sparsity allows more reliable signal detection performance and eventually could even improve the BER performance of the whole system without the reduction of the total bit per channel use (bpcu).

\begin{figure}[!tp]
     \centering
     \includegraphics[width=8.8cm, keepaspectratio]
     {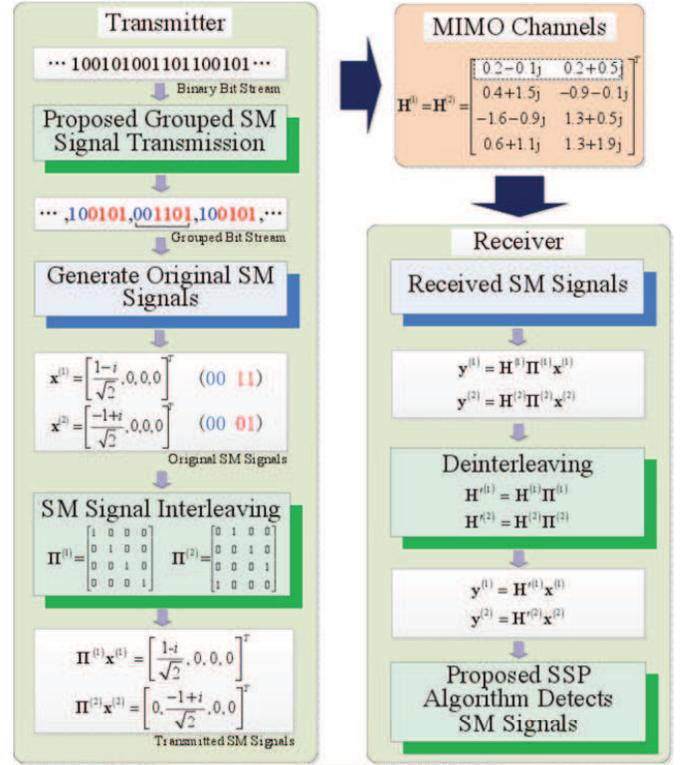}
    \caption{Illustration of the proposed SCS-based signal detector, where $N_t=4$, $N_r=2$, $N_a=1$, $G=2$, and QPSK are considered. The used spatial and signal constellation symbols can be illustrated in Fig. \ref{3D}. Note that the white dot bock in MIMO channels denotes the deep channel fading.}
     \label{fig:explain}
\end{figure}

On the other hand, due to the temporal channel correlation, channels in several consecutive time slots can be considered to be quasi-static, i.e., ${{\bf{H}}^{(1)}} = {{\bf{H}}^{(2)}} = \cdots  = {{\bf{H}}^{(G)}}$, where ${{\bf{H}}^{(t)}}$ for $1\le t \le G$ is the channel associated with the $t$th SM signal in the group. This implies that if channels used for SM fall into the deep fading, such deep fading usually remains unchanged during $G$ time slots, and the corresponding signal detection performance will be poor.
To solve this issue, we further propose the SM signal interleaving at the transmitter. Specifically, after the original SM signals ${\bf{x}}^{(t)}$'s are generated, the actually transmitted signals are given by ${{\bm{\Pi}} ^{\left( t \right)}} {{{\bf{x}}^{\left( t \right)}}} $'s, where each column and row of ${{\bm{\Pi}} ^{\left( t \right)}}\in \mathbb{C}^{N_t \times N_t}$ only has one non-zero element with the value of one, and ${{\bm{\Pi}} ^{\left( t \right)}}$ can permutate the entries in ${\bf{x}}^{(t)}$. We consider that ${\bm{\Pi}}^{(t)}$'s for $1\le t \le G$ are different in different time slots, and they are predefined and known by both the transmitter and receiver. In this way, the active antennas vary in different time slots from the same transmission group even though ${\bf{x}}^{(t)}$'s share the common spatial constellation symbol. Hence, the channel diversity can be appropriately exploited to improve the signal detection at the receiver. Note that the object of the SM signal interleaving is the spatial constellation symbol after constellation mapping, which is different from the widely used bit-interleaving whose object is the bit stream before constellation mapping~\cite{interleaving}.
In Section IV-B, such diversity gain will be further~discussed.

\subsection{SCS-Based Signal Detector at the Receiver}
At the receiver, the received signal in the $t$th time slot can be expressed as
  \begin{equation}
\begin{array}{l}
{{\bf{y}}^{\left( t \right)}} ={{\bf{H}}^{\left( t \right)}}{{\bm \Pi} ^{\left( t \right)}} {{{\bf{x}}^{\left( t \right)}}}  + {{\bf{w}}^{\left( t \right)}} \\
~~~~~= {{{\bf{H'}}}^{\left( t \right)}}{{\bf{x}}^{\left( t \right)}} + {{\bf{w}}^{\left( t \right)}},1 \le t \le G,
\label{CS_model0}
\end{array}
\end{equation}
where ${\bf{H}}'^{(t)} = {\bf{H}}^{(t)}{\bm{\Pi}}^{(t)}$ can be considered as the deinterleaving processing.

From (\ref{CS_model0}), we observe that ${{\bf{x}}^{\left( t \right)}}$'s share the structured sparsity due to the grouped transmission scheme at the transmitter, but they have different non-zero values due to the mutually independent signal constellation symbols. According to the theory of SCS, the structured sparsity of ${{\bf{x}}^{\left( t \right)}}$'s can be exploited to improve the signal detection performance compared with the conventional CS-based signal detectors~\cite{STR_CS}.
Under the framework of SCS theory, the solution to (\ref{CS_model0}) can be achieved by solving the following optimization problem:
\begin{equation}
\begin{array}{l}\label{opti} 
\!\!\! \mathop {\min }\limits_{{\rm{supp}}({{\bf{x}}^{\left(t \right)}}) \in {\mathbb{A}} } {\left( {\sum\limits_{t = 1}^G {\left\| {{{\bf{x}}^{\left( t \right)}}} \right\|_p^q} } \right)^{1/q}},\\~ {\rm s.t.} ~ {{\bf{y}}^{\left( t\right)}}\! =\! {{\bf{H'}}^{\left( t \right)}}{{\bf{x}}^{\left( t \right)}},~{\rm{supp}}\left( {{{\bf{x}}^{\left( t \right)}}} \right) = {\rm{supp}}\left( {{{\bf{x}}^{\left( 1\right)}}} \right), 1 \le t \le G.
\end{array}
\end{equation}
In this paper, based on the classical subspace pursuit (SP) algorithm \cite{SP}, we propose an SSP algorithm by utilizing the structured sparsity to solve the optimization problem~(\ref{opti}) in a greedy way, where $p=0$ and $q=2$ are advocated \cite{STR_CS}.

\begin{algorithm}[tp]
\begin{small}
\renewcommand{\algorithmicrequire}{\textbf{Input:}}
\renewcommand\algorithmicensure {\textbf{Output:} }
\caption{Proposed SSP Algorithm.}
\label{alg:Framwork} 
\begin{algorithmic}[1]
\REQUIRE
Received signal ${\bf{ y} }^{(t)}$, the channel matrix ${\bf{ H'}}^{(t)}$, and the number of active antennas $N_a$, where $1\le t \le G$.
\ENSURE
Estimated SM signal $ {{\hat {\bf{x}}^{\left( t \right)}}}$ for $1\le t \le G$. \\
\STATE $\Omega^0  = \emptyset$;~~~~~~~~~~~~~~~~~~~~~~~~~~~~~~${\kern 1.6pt}$~~\%~Empty $\Omega^0$ as $\emptyset$, and $\Omega^k$ is the estimated
 support set in the $k$th iteration.
\STATE ${\bf{r}}^{(t)}  = {\bf{y} }^{(t)}$, $\forall  t $; ~~~~~~~~~~~~~~~~~~~~~${\kern 3.9pt}$ \%~Initial residual, ${\bf{r}}^{(t)}$ is the residual associated with the $t$th SM signal.
\STATE $k= 1$;  ~~~~~~~~~~~~~~~~~~~~~~~~~~~~~~~~${\kern 4.4pt}$\%~$k$ is the iteration index.
\WHILE{$k\le N_a $}
\STATE ${\bf{a}}^{(t)} =\left( { {\bf{H'}} ^{\left( t \right)}} \right)^{*}{{\bf{r}}^{(t)}}$, $\forall  t $; ~~~~~~~~${\kern 1.9pt}$\%~Correlation 
\STATE $\Gamma \! =\! \arg \mathop {{\rm{max}}}\limits_{\tilde \Gamma }\!\! \left\{ {\sum\limits_{t = 1}^G {\left\| {{\bf{a}}_{\tilde \Gamma }^{\left( t \right)}} \right\|_2^2} ,\tilde \Gamma  \in \mathbb{A},\left| {\tilde \Gamma } \right|{\rm{ = }}\min \left\{ {2{N_a},{N_r}} \right\}{\rm{if}}~k = 1} \right.$\\
$\left. {{\rm{or}}\left| {\tilde \Gamma } \right|{\rm{ = }}\min \left\{ {{N_a},{N_r} - {N_a}} \right\}{\rm{if}}~k > 1} \right\}$;~~\%~Estimate potential supports.

\STATE ${\Xi  }={\Omega ^{k - 1}} \cup {\Gamma}$;~~~~~~~~~~~~~~~~~~~~${\kern 1.1pt}$\%~Merge estimated supports in the previous and current iteration.
\STATE ${\bf{b}}_\Xi ^{\left( t \right)} = \left( {{\bf{H}'}_\Xi ^{\left( t \right)}} \right)^{\dag }{{\bf{y}}^{\left( t \right)}}$, $\forall  t $;~~~~~~~~~${\kern 0pt}$\%~Least squares
\STATE ${\Omega ^k} = \arg \mathop {{\rm{max}}}\limits_{{\tilde \Omega}} \left\{ {\sum\limits_{t = 1}^G {\left\| {{\bf{b}}_{\tilde \Omega}  ^{\left(t \right)}} \right\|_2^2} ,{{\tilde \Omega} \in {\mathbb{A}} }  ~~\text{and}~~ {\left| {\tilde \Omega} \right|{\rm{ = }}N_a} } \right\}$;
~~${\kern 3.6pt}$\%~Prune the estimated support set.
\STATE ${\bf{c}}_{\Omega ^k} ^{\left( t\right)} =\left( { {\bf{H}'}_{\Omega ^k} ^{\left(t \right)}} \right)^{\dag}{{\bf{y}}^{\left(t \right)}}$, $\forall  t $; ~~~~~~~~${\kern -0.5pt}$\%~Least squares
\STATE ${\bf{r}}^{(t)} = {\bf{ y} }^{(t)}- {{\bf{ H  '}}^{(t)} }{\bf{c}}^{(t)}$, $\forall  t $;~~~~~~~${\kern -0.9pt}$\%~Compute residual
\STATE $k= k+1$;
\ENDWHILE
\STATE ${\bf{\hat x}}_{{\rm{}}}^{\left( t \right)} = {\bf{c}}_{}^{\left( t \right)}$, $\forall  t $;
\end{algorithmic}
\end{small}
\end{algorithm}

The proposed SSP algorithm is described in \textbf{Algorithm 1}. Specifically, {\it Lines 1$\sim$3} perform the initialization. In the $k$th iteration, {\it Line 5} performs the correlation between the MIMO channels and the residual in the previous iteration; {\it Line 6} obtains the potential true indices according to {\it Line 5}; {\it Line 7} merges the estimated indices obtained in {\it Lines 8$\sim$9} in the previous iteration and the estimated indices in {\it Line 6} in the current iteration; after the least squares in {\it Line 8}, {\it Line 9} removes wrong indices and selects $N_a$ most likely indices; {\it Line 10} estimates SM signal according to $\Omega^k$; {\it Line 11} acquires the residue. The iteration stops when $k>N_a$
\footnote{For the classical SP algorithm, when the residual
of current iteration is not less than that of the last iteration, the iteration stops and
the estimation in the last iteration is used as the final output, which is different from the proposed stopping criterion with the fixed number of iteration. Simulation results confirm both stopping criteria are equivalent in most cases and they share the very similar performance for the proposed SSP algorithm. Besides, for real-time or delay sensitive applications, it is usually desirable to have the number of iteration be fixed or
bounded, so that the speed of decoding and power consumption is manageable.}.Compared with the classical SP algorithm which only reconstructs one sparse signal from one received signal, the proposed SSP algorithm can jointly recover multiple sparse signals with the structured sparsity but having different measurement matrices, where the structured sparsity of multiple sparse signals can be leveraged for improved signal detection performance. Therefore, the classical SP algorithm can be regarded as a special case of the proposed SSP algorithm when $G=1$, and more details will be further discussed in Section \ref{SCS_CS}. Another difference should be pointed out that in the steps of {\it{Line 6}} and {\it 9} in \textbf{Algorithm 1}, the selected support set should belong to the predefined spatial constellation set $\mathbb{A}$ for enhanced signal detection performance. However, the classical SP algorithm and existing CS-based signal detectors do not exploit this priori information of the expected support set \cite{{CS_CL1},{CS_CL2}}.
By using the proposed SSP algorithm, we can acquire the estimation of the spatial constellation symbol according to ${\rm{supp}}\left( {{\bf{\hat x}}_{}^{\left( t \right)}} \right)$'s and the rough estimation of signal constellation symbols. By searching for the minimum Euclidean distance between the rough estimation of signal constellation symbols and legitimate constellation symbols, we can finally estimate signal constellation symbols.

\section{Performance Analysis}
In this section, we will provide the performance analysis from four aspects as follows.
\subsection{Comparison Between SCS-Based Signal Detector and CS-Based Signal Detectors}\label{SCS_CS}
Typically, existing CS-based signal detectors utilize one received signal vector to recover one sparse SM signal vector, which is equivalent to solving the single measurement vector (SMV) problem in CS, i.e., ${\bf{y}} = {\bf{Hx}}+{\bf{w}} $.
If multiple sparse signals share the common support set and identical measurement matrix, i.e., $\left[ {{{\bf{y}}^{(1)}},{{\bf{y}}^{(2)}}, \cdots ,{{\bf{y}}^{(G)}}} \right] = {\bf{H}}\left[ {{{\bf{x}}^{(1)}},{{\bf{x}}^{(2)}}, \cdots ,{{\bf{x}}^{(G)}}} \right]+{\bf{w}} $, the reconstruction of $ {{\bf{x}}^{(t)}}$'s from ${{\bf{y}}^{(t)}}$'s for $1\le t \le G$ can be considered as the multiple measurement vectors (MMV) problem in SCS theory \cite{STR_CS}. The SCS theory has proven that with the same size of the measurement vector, the recovery performance of SCS algorithms is superior to that of conventional CS algorithms~\cite{STR_CS}. This implies that with the same number of receive antennas $N_r$, the proposed SCS-based signal detector can outperform conventional CS-based signal detectors.

Compared with the conventional MMV problem, our formulated problem (\ref{opti}) is to solve multiple sparse signals with the common support set but having different measurement matrices due to the proposed SM signal interleaving. Hence both conventional SMV problem and MMV problem can be considered as the special cases of our problem. If ${{\bm{\Pi}} ^{\left( t \right)}}$'s are identical, our problem (\ref{opti}) becomes the conventional MMV problem, and furthermore if $G=1$, it reduces to the SMV problem. Therefore, our formulated problem can be regarded as a generalized MMV (GMMV) problem.

%



\subsection{Performance Gain from SM Signal Interleaving}\label{interleaving}
We discuss the performance gain from the SM signal interleaving by comparing the detection probability of the proposed SSP algorithm with and without SM signal interleaving. Since CS algorithms are nonlinear, it is difficult to exactly provide the closed-form expression of the signal detection probability. Hence, we consider a simplified scenario with $N_a=1$ and uncorrelated Rayleigh-fading MIMO channels.
Let $m$ be the index of the active antenna, and for any given $l$, ${{{\bf{H}'}^{(t)}_l}}$'s for $1\le t \le G$ are mutually
independent\footnote{In our problem, measurement matrices ${{{\bf{H'}}}^{\left( t \right)}}$'s are not mutually independent, since ${{{\bf{H'}}}^{\left( t \right)}}$'s are generated by permuting the columns of ${{\bf{H}}^{(1)}} = {{\bf{H}}^{(2)}} = \cdots  =
{{\bf{H}}^{(G)}}$ with different permutation matrices. Fortunately, we can appropriately design ${{\bm{\Pi}} ^{\left( t \right)}}$'s to guarantee that ${{{\bf{H}'}^{(t)}_l}}$'s for a given $l$ are mutually independent, and the proposed SM signal interleaving enables us to approach the performance gain of the ideal case with mutually independent ${{{\bf{H'}}}^{\left( t \right)}}$'s, which will be verified in Section~V.},
where $1\le m, l \le N_t$. Based on these assumptions, the received signal is given by ${{\bf{y}}^{(t)}} = {\alpha ^{(t)}} {{{\bf{H}'}^{(t)}_m}}+{\bf{w}}^{(t)}$, for $1\le t \le G$, where ${\alpha ^{(t)}} \in \mathbb{B}$ denotes the signal constellation symbol carried by the active antenna in the $t$th time slot. To identify the active antenna, the proposed SSP algorithm relies on the correlation operation in \emph{Line 5} of\textbf{ Algorithm 1}, i.e.,
 \begin{small}
 \begin{equation}
\begin{array}{l}
\!\!\!\!{C_l} \triangleq  \sum\limits_{t = 1}^G {{{\left| {{{\left( {{{\bf{y}}^{(t)}}} \right)}^*}{\bf{H'}}_l^{\left( t \right)}} \right|}^2}} \!\!\! =\!\!\! \sum\limits_{t = 1}^G {{{\left| {{{\left( {{\alpha ^{(t)}}{\bf{H'}}_m^{\left( t \right)}} +{\bf{w}}^{(t)}\right)}^*}{\bf{H'}}_l^{\left( t \right)}} \right|}^2}}  \!\!\! =\!\!\! \sum\limits_{t = 1}^G {{{\left| {F_{m,l}^{(t)}} \right|}^2}},
\end{array}
\end{equation}
\end{small}where $F_{m,l}^{(t)} =  {{{\left( {{\alpha ^{(t)}}{\bf{H'}}_m^{\left( t \right)}} +{\bf{w}}^{(t)}\right)}^*}{\bf{H'}}_l^{\left( t \right)}} $ for $1\le l \le N_t$.
Due to large $N_r$ in practice, we have ${\mathop{\rm Re}\nolimits} \left\{ {F_{m,m}^{(t)}} \right\}\sim {\cal{N}}\left( {{\mu _1},\sigma _1^2} \right)$ with
${\mu _{\rm{1}}} =0$, 
$\sigma _1^2  
=  \frac{{(N_r^2 + N_r^{})\sigma _s^2}}{{2 - \delta \left( {M = 2} \right)}} + \frac{{N_r^{}\sigma _w^2}}{2}$,
and ${\mathop{\rm Im}\nolimits} \left\{ {F_{m,m}^{(t)}} \right\}\sim {\cal{N}}\left( {{\mu _2},\sigma _2^2} \right)$ with ${\mu _{\rm{2}}} = 0$, $\sigma _2^2 = \frac{{\left( {1 - \delta \left( {M = 2} \right)} \right)(N_r^2 + N_r^{})\sigma _s^2}}{2} + \frac{{N_r^{}\sigma _w^2}}{2}$
according to central limit theorem \cite{Sig_det}.
Similarly, both ${\mathop{\rm Re}\nolimits} \left\{ {F_{m,l}^{(t)}} \right\}$ and ${\mathop{\rm Im}\nolimits} \left\{ {F_{m,l}^{(t)}} \right\}$ follow the distribution $ {\cal{N}}\left( {{\mu _3},\sigma _3^2} \right)$ with $ l\neq m$, ${\mu _{\rm{3}}} =0$,
$ \sigma _3^2 =  \frac{  {N_r^{}\sigma _s^2}}{{2}}+\frac{  {N_r^{}\sigma _w^2}}{{2^{}}}$. The associated proof will be provided in Appendix.
Note that 
$\sigma _s^2 = {\rm{Tr}}\left\{ {E\left\{ {{{\bf{x}}^{\left( t \right)}}{{\left( {{{\bf{x}}^{\left( t \right)}}} \right)}^T}} \right\}} \right\}$, and ${\mathop{\rm Re}\nolimits} \left\{ {F_{m,l}^{(t)}} \right\}$ and ${\mathop{\rm Im}\nolimits} \left\{ {F_{m,l}^{(t)}} \right\}$ $\forall l$ are mutually independent.
Moreover, we can have ${C_m} \sim \sigma _2^2\chi _{{G}}^2 + \sigma _1^2\chi _{{G}}^2$ and ${C_l} \sim \sigma _3^2\chi _{2G{}}^2$ with $l\neq m$, where $\chi _n^2$ is the central chi-squared distribution with the degrees of freedom $n$ \cite{Sig_det}.
Since \textbf{Algorithm 1} only has one iteration and $\left| { \Gamma } \right|=\left| {\Xi  } \right|=2$ in the iteration for $N_a=1$, we consider ${P_{{\rm{GMMV}}}}\left( {{C_m} - C_l^{[2]} > 0|l \ne m} \right)$ as the correct active antenna detection probability, where ${C_l^{[1]}}>{C_l^{[2]}}> \cdots > {C_l^{[N_t-N_a]}}$ with $l\neq m$ are sequential statistics. The probability density functions (PDFs) of $C_m$ and $C_l^{}$ with $l\neq m$ are denoted by $f_1(x)$ and $f_2(x)$, respectively. The PDF of ${{C_l}}^{[2]}$ with $l \neq m$ is
$f_2^{[2]}\left( x \right) = \frac{{\left( {{N_t} - {N_a}} \right)!}}{{\left( {{N_t} - {N_a} - 2} \right)!}}{\left( {{F_2}\left( x \right)} \right)^{{N_t} - {N_a} - 2}}{\left( {1 - {F_2}\left( x \right)} \right)^{}}{f_2}\left( x \right)$,
where ${F_2}\left( x \right)$ is the cumulative density function of $f_2(x)$. In this way, we have
  \begin{small}
 \begin{equation}
\begin{array}{l}
\!\!\!\!{P_{{\rm{GMMV}}}}\!\! \left(\! {{C_m}\!\! -\!\! C_l^{[2]} \!> \! 0|l \!\ne\! m} \right) \!\!=\!\! \int_0^\infty \!\!  {\int_{ - \infty }^\infty  \!\! {f\left( x \right){f_{2}^{[2]}}\!\left( {x\! -\! z} \right)dxdz} }.
\end{array}
\end{equation}
\end{small}

For the conventional MMV problem with identical channel matrices, similar to the previous analysis, we have ${C_m} \sim G\sigma _2^2\chi _{{1}}^2 + G\sigma _1^2\chi _{{1}}^2$ and ${C_l} \sim G\sigma _3^2\chi _{2{}}^2$ with $l\neq m$. Similarly, we can also get ${P_{{\rm{MMV}}}}\left( {{C_m} - C_l^{[2]} > 0|l \ne m} \right)$.

To intuitively compare the signal detection probability, we compare ${P_{{\rm{MMV}}}}\left( {\left. {{C_m} - {C_l} > 0} \right|l \ne m} \right)$ and ${P_{{\rm{GMMV}}}}\left( {\left. {{C_m} - {C_l} > 0} \right|l \ne m} \right)$ when $\sigma _s^2/\sigma _w^2 \to \infty $ and $G$ are sufficient large. In this case, $C_m-C_l$ can be approximated to the Gaussian distribution ${\cal{N}}\left( {{\mu _4},\sigma _4^2} \right)$ with ${\mu _4} = G\left( {\mu _1^2 + \mu _2^2 - 2\mu _3^2 + \sigma _1^2 + \sigma _2^2 - 2\sigma _3^2} \right)$, $\sigma _4^2 = G\sum\nolimits_{i = 1}^3 {2\sigma _i^4 + 4\mu _i^2\sigma _i^2} $. In this way, we can obtain that ${P_{{\rm{GMMV}}}}\left( {\left. {{C_m} - {C_l} > 0} \right|l \ne m} \right) \approx Q(-{\mu _4}/\sigma _4^{})$, where $Q$-function is the tail probability of the standard normal distribution \cite{Sig_det}. By contrast, for conventional MMV case, we can obtain that ${P_{{\rm{MMV}}}}\left( {\left. {{C_m} - {C_l} > 0} \right|l \ne m} \right)  \approx Q(-{\mu _4}/(\sqrt{G}\sigma _4^{}))$.
Clearly, ${P_{{\rm{MMV}}}}$ is larger than ${P_{{\rm{GMMV}}}}$ due to ${\mu _4} >0$ and $G>1$, which implies that an appropriate SM signal interleaving will lead to the improved signal detection performance.

To achieve the goal that ${{{\bf{H}'}^{(t)}_l}}$'s, $\forall l$, are mutually independent as much as possible, we consider the pseudo-random permutation matrix ${{\bm \Pi }^{\left( t \right)}}$, which can be predefined and known by both the BS and user.
In Section \ref{sim}, simulation results confirm the good performance gain of the channel diversity from SM signal interleaving, whose performance gain approaches that of the case of mutually independent channel matrices in the same group.

\subsection{Spectral Efficiency}
The proposed scheme has the spectrum efficiency ${N_a}{\log _2}M + {\rm{lo}}{{\rm{g}}_2}\left| \mathbb{A} \right|/G$ bpcu due to the grouped transmission scheme, which is slightly smaller than conventional SM-MIMO systems with ${N_a}{\log _2}M + {\rm{lo}}{{\rm{g}}_2}\left| \mathbb{A} \right|$ bpcu. In SM-MIMO, the detection of spatial constellation symbols is essential since it is the prerequisite of the following detection of signal constellation symbols. Conventional low-complexity signal detectors such as CS-based signal detectors perform poorly in massive SM-MIMO systems, and thus the signal constellation symbol is usually limited to low-order modulation to guarantee the reliable signal detection. However, the detection of spatial constellation symbol can be improved by the proposed scheme, so higher-order modulation of signal constellation symbol can be used to achieve the same or even higher bpcu. Besides, the spectrum efficiency can be further increased by using a large number of low-cost transmit antennas to expand the degree of spatial freedom in massive SM-MIMO. Finally, simulation results in the following section will show that even with small $G=2$, the proposed SCS-based signal detector can achieve much better BER improvement even with higher total bpcu than the conventional signal detectors.
\begin{figure}[!tp]
     \centering
     \includegraphics[width=8.5cm, keepaspectratio]
     {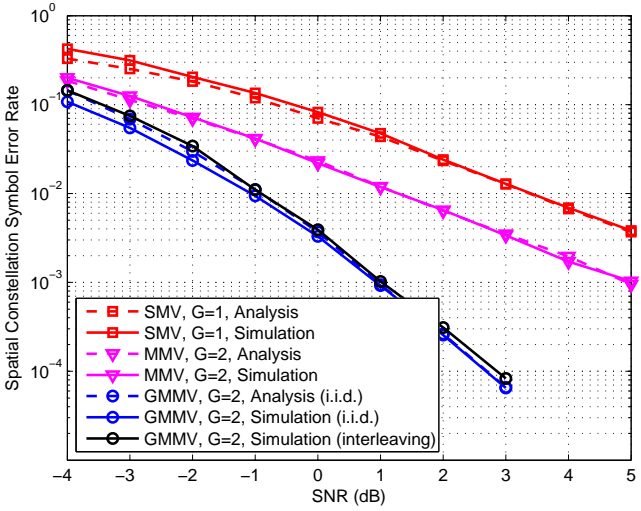}
    \caption{Comparison of the simulated and analytical SCSER of the SCS-based signal detector in different cases over uncorrelated Rayleigh-fading MIMO channels, where $N_t=64$, $N_r=16$, $N_a=1$, and 8-PSK are considered. }
     \label{fig:fig1}
\end{figure}
\subsection{Computational Complexity}
The optimal ML signal detector has the computational complexity of ${\cal{O}}({M^{{N_a}}}{2^{\left\lfloor {{\rm{log_2}}\binom{N_t}{N_a}} \right\rfloor }})$, which is prohibitively high when $N_a$, $N_t$, and/or $M$ become large. The conventional signal detectors \cite{{LS_SM},{CS_CL1},{SD}} have the complexity of ${\cal{O}}(N_t^3)$ as mentioned in Section \ref{model}, which implies that their complexity is still high in massive SM-MIMO systems with large $N_t$.
By contrast, for the proposed SCS-based signal detector, the main computational burden comes from the step of least squares with the computational complexity of ${\cal{O}}(G(2N_r N_a^2 + N_a^3))$ \cite{LS}, or equivalently ${\cal{O}}(2N_r N_a^2 + N_a^3)$ per SM signal in each time slot. This indicates that the proposed SCS-based signal detector enjoys the same order of complexity with the CS-based signal detector \cite{{CS_CL2}}.

\section{Simulation Results}\label{sim}
A simulation study was carried out to compare the performance of the proposed SCS-based signal detector with that of the conventional LMMSE-based signal detector \cite{LS_SM} and the CS-based signal detector \cite{{CS_CL1}}. The performance of the optimal ML detector \cite{ML2} is also provided as the benchmark for comparison.


Fig. \ref{fig:fig1} compares the simulated and analytical spatial constellation symbol error rate (SCSER) of the SCS-based signal detector in different cases over uncorrelated Rayleigh-fading MIMO channels, where $N_t=64$, $N_r=16$, $N_a=1$, and 8-PSK are considered. For the GMMV case, ``i.i.d." denotes the case that ${\bf{H'}}_{}^{\left( t \right)} = {\bf{H}}_{}^{\left( t \right)}$ for $1\le t \le G$ and ${\bf{H}}_{}^{\left( t \right)}$'s are independently generated, while ``interleaving" denotes the case that ${{\bf{H}}^{(1)}} = {{\bf{H}}^{(2)}} = \cdots  = {{\bf{H}}^{(G)}}$ and ${\bf{H}}'^{(t)} = {\bf{H}}^{(t)}{\bm{\Pi}}^{(t)}$ with different permutation matrices ${\bm{\Pi}}^{(t)}$'s. From Fig. \ref{fig:fig1}, we can find that the analytical SCSER derived in Section \ref{interleaving} have the good tightness with the simulation results.
In addition, the proposed SCS-based signal detector outperforms the conventional CS-based signal detector, since the structured sparsity of multiple sparse SM signals is exploited for the improved SCSER. Moreover, since the channel diversity can be exploited to further improve the SCSER, the SCS-based signal detector with mutually independent channel matrices (GMMV) is superior to that with identical channel matrices (MMV) by more than 4~dB if the SCSER of $10^{-3}$ is considered. Finally, the performance of the SCS-based signal detector with SM signal interleaving approaches that with mutually independent channel matrices, which indicates that the proposed SM signal interleaving can fully exploit the channel diversity.

\begin{figure}[!tp]
     \centering
     \includegraphics[width=8.5cm, keepaspectratio]
     {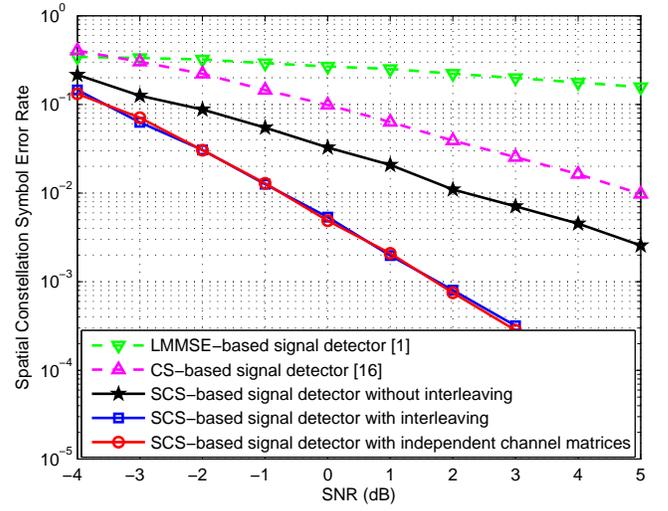}
    \caption{SCSER of different signal detectors over correlated Rayleigh-fading MIMO channels, where $r_t=r_r=0.4$, $N_t=64$, $N_r=16$, $N_a=1$, and 8-PSK are considered.}
     \label{fig:fig2}
\end{figure}

\begin{figure}[!tp]
     \centering
     \includegraphics[width=8.5cm, keepaspectratio]
     {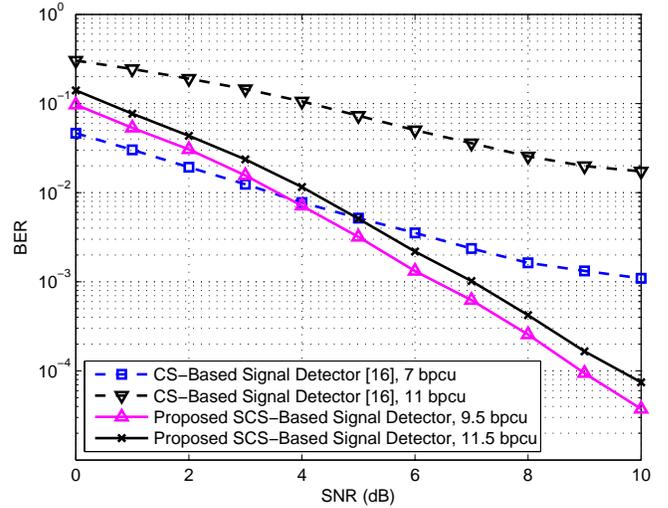}
    \caption{BER comparison between the traditional CS-based signal detector and the proposed SCS-based signal detector over correlated Rayleigh-fading MIMO channels, where $r_t=r_r=0.4$ and $N_r=16$.}
     \label{fig:fig3}
\end{figure}

Fig. \ref{fig:fig2} provides SCSER comparison of different signal detectors over correlated Rayleigh-fading MIMO channels, where both the channel correlation coefficients at the transmitter and receiver are $r_t=r_r=0.4$ \cite{CE_SM}, $N_t=64$, $N_r=16$, $N_a=1$, and 8-PSK are considered. The conventional LMMSE-based signal detector works poorly due to $N_r\ll N_t$. The SCS-based signal detector with interleaving outperforms the conventional CS-based signal detector and SCS-based signal detector without interleaving. Moreover, it has the similar performance with that with mutually independent channel matrices (i.e., ${\bf{H'}}_{}^{\left( t \right)} = {\bf{H}}_{}^{\left( t \right)}$ for $1\le t \le G$ and ${\bf{H}}_{}^{\left( t \right)}$'s are independently generated), which indicates the good performance gain of the channel diversity from interleaving even in correlated MIMO channels.

Fig. \ref{fig:fig3} provides the BER performance comparison of the existing CS-based signal detector and the proposed SCS-based signal detector  with interleaving over correlated Rayleigh-fading MIMO channels with $r_t=r_r=0.4$ and $N_r=16$. The existing scheme adopts two transmission modes: 1) $N_t=64$, $N_a=1$, BPSK with 7 bpcu and 2) $N_t=65$, $N_a=2$, no signal constellation symbol with 11 bpcu. In contrast, the SCS-based signal detector with $N_t=65$, $N_a=2$ and $G=2$ adopts QPSK and 8-PSK, respectively, and the corresponding data rates are 9.5 bpcu and 11.5 bpcu. From Fig. \ref{fig:fig3}, it can be observed that the proposed SCS-based signal detector with even higher bpcu achieves better BER performance than the conventional CS-based signal detector. For example, when BER of $10^{-3}$ is considered, the proposed SCS-based signal detector with 9.5 bpcu outperforms the conventional CS-based signal detector with 7 bpcu by about 2 dB.

\begin{figure}[!tp]
     \centering
     \includegraphics[width=8.5cm, keepaspectratio]
     {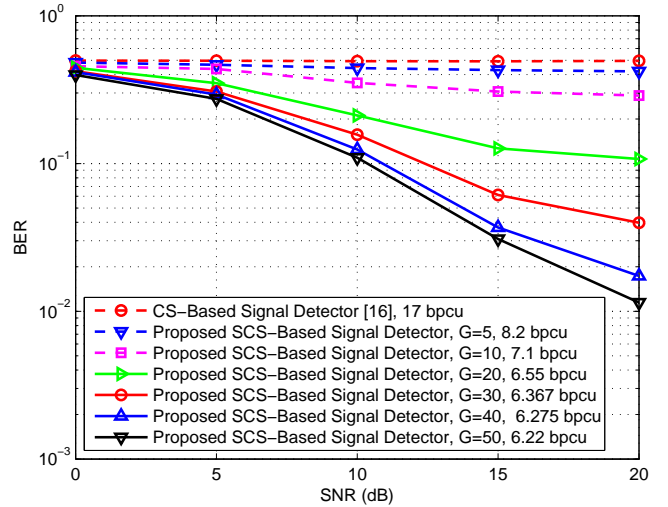}
    \caption{BER performance between the conventional CS-based signal detector and the proposed SCS-based signal detector with different $G$'s, where $r_t=r_r=0.4$, $N_t=65$, $N_r=3$, $N_a=2$, 8-PSK are considered.}
     \label{fig:fig4}
\end{figure}

In Fig. \ref{fig:fig4}, we compare the BER performance between the conventional CS-based signal detector and the proposed SCS-based signal detector with interleaving over correlated Rayleigh-fading MIMO channels with $r_t=r_r=0.4$ and different $G$'s, where an extreme case that $N_t=65$, $N_r=3$, $N_a=2$, and 8-PSK is considered\footnote{The proposed scheme may require a relatively large number of receive antennas to achieve its competitive advantage, and it can be applied for the mobile stations such as laptop or tablet, where the relatively large number of antennas may be possible at the mobile stations, especially when the higher working frequency of 3$\sim$6~GHz is considered.}. The conventional CS-based signal detector works poorly, since the limitation $N_r \ge 2N_a+1$ is required by conventional CS theory \cite{STR_CS}. By contrast, the BER performance of the proposed SCS-based signal detector improves when $G$ becomes large, since the SCS theory can relax the limitation as $N_r \ge N_a+1$ \cite{STR_CS}.
\begin{figure}[!tp]
     \centering
     \includegraphics[width=8.5cm, keepaspectratio]
     {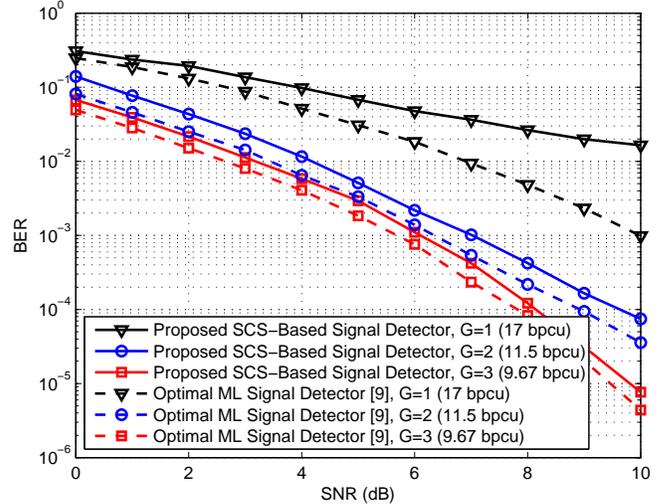}
    \caption{BER performance comparison between the proposed SCS-based signal detector and the optimal ML signal detector, where $r_t=r_r=0.4$, $N_t=65$, $N_r=16$, $N_a=2$, and 8-PSK are considered.}
     \label{fig:fig5}
\end{figure}

Fig. \ref{fig:fig5} compares the performance of the proposed SCS-based signal detector with interleaving and the optimal ML signal detector, where $r_t=r_r=0.4$, $N_t=65$, $N_r=16$, $N_a=2$, and 8-PSK are considered. 
From Fig. \ref{fig:fig5}, we can find that with the increasing $G$, the BER performance gap between the SCS-based signal detector and the optimal ML signal detector becomes smaller. When $G \ge 2$, the SCS-based signal detector approaches the optimal ML signal detector with a small performance loss. For example, if the BER of $10^{-4}$ is considered, the performance gap between the SCS-based signal detector with $G = 3$ and the optimal ML detector is less than 0.2~dB. Thus, the near-optimal performance of the proposed SCS-based signal detector can be verified.
\section{Conclusions}
This paper has proposed a near-optimal SCS-based signal detector with low complexity for the emerging massive SM-MIMO. 
First, the grouped transmission scheme can introduce the desired structured sparsity of multiple SM signals in the same transmission group for improved signal detection performance. Second, the SSP algorithm can jointly detect multiple SM signals with low complexity. Third, by using SM signal interleaving, we can fully exploit the channel diversity to further improve the signal detection performance, and the performance gain from SM signal interleaving can approach that of the ideal case of mutually independent channel matrices in the same transmission group. Besides, we have quantified the performance gain from SM signal interleaving. Simulation results have confirmed that the proposed low-complexity SCS-based signal detector outperforms conventional signal detectors with near-optimal performance.
\appendix\label{app}
We will investigate the distribution of $F_{m,l}^{} =   {{{\left( {{\alpha ^{}}{\bf{H'}}_m^{}} +{\bf{w}}^{}\right)}^*}{\bf{H'}}_l^{}} $, where the superscript $(t)$ is omitted for simplicity. Specifically, let $h_{m,n}$, $h_{l,n}$, and $w_{n}$ denote the $n$th element of the vector
${\bf{H'}}_m^{}$, ${\bf{H'}}_l^{}$, and ${\bf{w}}^{}$, respectively. In this way, we have
 $F_{m,l}^{} = \alpha \sum\nolimits_{n = 1}^{{N_r}} {h_{m,n}^*{h_{l,n}}}  + {\sum\nolimits_{n = 1}^{{N_r}} {w_n^*{h_{l,n}}} ^{}}$.
Since $F_{m,l}^{}$ can be expressed as the summation of multiple mutually independent random variables, $F_{m,l}^{}$ approximately follows the Gaussian distribution, when $N_r$ is sufficiently large according to central limit theorem.
 Obviously, we have $E\left\{ {F_{m,l}^{}} \right\}=0$, since ${\mathop{\rm Re}\nolimits} \left\{ {{h_{l,n}}} \right\}=h_{l,n}^r$, ${\mathop{\rm Im}\nolimits} \left\{ {{h_{l,n}}} \right\}=h_{l,n}^i$, ${\mathop{\rm Re}\nolimits} \left\{ {{h_{m,n}}} \right\}=h_{m,n}^r$, and ${\mathop{\rm Im}\nolimits} \left\{ {{h_{m,n}}} \right\}=h_{m,n}^i$ follow ${\cal{N}}(0,0.5)$, both ${\mathop{\rm Re}\nolimits} \left\{ {{w_{n}}} \right\}=w_{n}^r$ and ${\mathop{\rm Im}\nolimits} \left\{ {{w_{n}}} \right\}=w_{n}^i$ follow ${\cal{N}}(0,\sigma_w^2/2 )$, and $E\left\{ \alpha  \right\} = 0$.
 For the case of $m=l$, we have
  \begin{equation}\label{equ:A1}
   \begin{footnotesize}
\begin{array}{l}
F_{m,m}^{} = \alpha \sum\nolimits_{n = 1}^{{N_r}} {{{\left| {{h_{m,n}}} \right|}^2}}  + \sum\nolimits_{n = 1}^{{N_r}} {w_n^*{h_{m,n}}} \\
 = \sum\nolimits_{n = 1}^{{N_r}} {{\rm{Re}}\left\{ \alpha  \right\}\left( {{{(h_{m,n}^r)}^2} + {{(h_{m,n}^i)}^2}} \right) + h_{m,n}^rw_n^r + h_{m,n}^iw_n^i} \\
+ i\sum\nolimits_{n = 1}^{{N_r}} {{\rm{Im}}\left\{ \alpha  \right\}\left( {{{(h_{m,n}^r)}^2} + {{(h_{m,n}^i)}^2}} \right) - h_{m,n}^rw_n^i + h_{m,n}^iw_n^r}.
\end{array}
 \end{footnotesize}
  \end{equation}
  Furthermore, we can have
 \begin{equation}\label{equ:A2}
   \begin{footnotesize}
\begin{array}{l}
E\left\{ {{\mathop{\rm Re}\nolimits} {{\left\{ {F_{m,m}^{}} \right\}}^2}} \right\}\\ = E\left\{ {{{\left( {\sum\limits_{n = 1}^{{N_r}} {{\rm{Re}}\left\{ \alpha  \right\}\left( {{{(h_{m,n}^r)}^2} + {{(h_{m,n}^i)}^2}} \right) + h_{m,n}^rw_n^r + h_{m,n}^iw_n^i} } \right)}^2}} \right\}\\
 = E\left\{ {\sum\limits_{n = 1}^{{N_r}} {{\rm{Re}}{{\left\{ \alpha  \right\}}^2}\left( {{{(h_{m,n}^r)}^4} + {{(h_{m,n}^i)}^4}} \right) + {{(h_{m,n}^rw_n^r)}^2} + {{(h_{m,n}^iw_n^i)}^2}} } \right\}\\
 + E\left\{ {{\rm{Re}}{{\left\{ \alpha  \right\}}^2}\sum\limits_{{n_1} = 1}^{{N_r}} {\sum\limits_{{n_2} = 1}^{{N_r}} {\left( {{{(h_{m,{n_1}}^rh_{m,{n_2}}^i)}^2} + {{(h_{m,{n_1}}^ih_{m,{n_2}}^r)}^2}} \right)} } } \right\}\\
 + E\left\{ {{\rm{Re}}{{\left\{ \alpha  \right\}}^2}\sum\limits_{{n_1} = 1}^{{N_r}} {\sum\limits_{{n_2} = 1,{n_2} \ne {n_1}}^{{N_r}} {\left( {{{(h_{m,{n_1}}^rh_{m,{n_2}}^r)}^2} + {{(h_{m,{n_1}}^ih_{m,{n_2}}^i)}^2}} \right)} } } \right\}\\
 + E\left\{ {{\rm{Re}}\left\{ \alpha  \right\}\sum\limits_{{n_1} = 1}^{{N_r}} {\sum\limits_{{n_2} = 1}^{{N_r}} {\left( {{{(h_{m,{n_1}}^r)}^2} + {{(h_{m,{n_1}}^i)}^2}} \right)} } } \right.\\
 \times \left. {(h_{m,{n_2}}^rw_{{n_2}}^r + h_{m,{n_2}}^iw_{{n_2}}^i) + h_{m,{n_1}}^rw_{{n_1}}^rh_{m,{n_2}}^iw_{{n_2}}^i} \right\}\\
 = \frac{{3E\left\{ {{\rm{Re}}{{\left\{ \alpha  \right\}}^2}} \right\}{N_r}}}{2} + \frac{{{N_r}\sigma _w^2}}{2} + \frac{{E\left\{ {{\rm{Re}}{{\left\{ \alpha  \right\}}^2}} \right\}N_r^2}}{2} + \frac{{E\left\{ {{\rm{Re}}{{\left\{ \alpha  \right\}}^2}} \right\}N_r^{}\left( {N_r^{} - 1} \right)}}{2} + 0\\
 = {{E\left\{ {{\rm{Re}}{{\left\{ \alpha  \right\}}^2}} \right\}(N_r^2 + {N_r})}}{} + \frac{{{N_r}\sigma _w^2}}{2}.
\end{array}
   \end{footnotesize}
  \end{equation}
  Similarly, we have
   \begin{equation}
 E\left\{ {{\mathop{\rm Im}\nolimits} {{\left\{ {F_{m,m}^{}} \right\}}^2}} \right\} = {{E\left\{ {{\rm{Im}}{{\left\{ \alpha  \right\}}^2}} \right\}(N_r^2 + {N_r})}} + {{{N_r}\sigma _w^2}}/2.
   \end{equation}
For BPSK with $M=2$, we have $E{\left\{ {{\rm{Re}}{{\left\{ \alpha  \right\}}^2}} \right\}}=\sigma _s^2$ and $E{\left\{ {{\rm{Im}}{{\left\{ \alpha  \right\}}^2}} \right\}}=0$, while for the higher constellation modulation with $M>2$ (e.g., $M$-QAM or $M$-PSK), we have $E{\left\{ {{\rm{Re}}{{\left\{ \alpha  \right\}}^2}} \right\}}=\sigma _s^2/2$ and $E{\left\{ {{\rm{Im}}{{\left\{ \alpha  \right\}}^2}} \right\}}=\sigma _s^2/2$.
\\
For the case of $m\neq l$, the associated result can be derived similarly to (\ref{equ:A1}) and (\ref{equ:A2}).


\end{document}